\documentclass{revtex4}
\usepackage{amsmath}
\usepackage{amssymb}
\usepackage{graphicx}

\begin{document}

\title{Phantom fields: bounce solutions in the early Universe and S-branes}
\author{Vladimir Dzhunushaliev
\footnote{Senior Associate of the Abdus Salam ICTP}}
\email{dzhun@krsu.edu.kg}
\affiliation{Dept. Phys. and Microel. Engineer., Kyrgyz-Russian Slavic University, Bishkek, Kievskaya Str. 44, 720021, Kyrgyz Republic}

\author{Vladimir Folomeev}
\email{vfolomeev@mail.ru} \affiliation{Institute of Physics of NAS
KR, 265 a, Chui str., Bishkek, 720071,  Kyrgyz Republic}

\author{Kairat Myrzakulov}
\affiliation{Institute of Physics and
Technology, 050032, Almaty, Kazakhstan}

\author{Ratbay Myrzakulov}\email{cnlpmyra@satsun.sci.kz}
\affiliation{Institute of Physics and
Technology, 050032, Almaty, Kazakhstan}

\begin{abstract}
The cosmological model with two phantom scalar fields with the
special choice of field's potential is considered. The obtained
regular solution describes a bounce with a subsequent transition
to the de Sitter stage of the expansion of the Universe. This
solution also could be interpreted as a S-brane solution.

\end{abstract}

\maketitle

\section{Introduction}

Recent observations of Type Ia supernovae show an accelerating
Universe at the present  epoch~\cite{Riess, Star}. Such accelerated
expansion is caused by presence of some  antigravitating substance
which act as repulsive force in the Universe. The immediate
observation of this substance is sufficiently hard due to its weak
interaction with electromagnetic radiation. In this connection this
substance was called dark energy (DE). The  true nature of DE
remains a puzzle till now.  The simplest model is a description of
DE as a  constant $\Lambda$-term (the so-called  $\Lambda$CDM model,
see~\cite{Star}, for example). However, at the present time it is rather
difficult to find physically proved explanation of a value and
nature of the $\Lambda$-term.

In this connection, consideration of dynamical models of DE with the varying equation of state seems to be more natural.
In this case the effective equation of state of DE corresponds to $w<-1/3$, while for the  $\Lambda$CDM model $w =-1$.
All the models laying in these
limits on $w$ can be  divided into two big categories: 1) the source of DE is different scalar fields (quintessence models,
Chaplygin gas, etc. \cite{Ratra, Zlatev, Kam}); 2) presence of DE is a consequence of modification of the Einstein gravitational
equations by addition of terms depending on scalar curvature, invariants of the Ricci and Riemann tensors, etc.
\cite{Turner,Capoz,fol,Noj}. Consideration of the evolution of the present Universe within the
framework of these models allows to achieve the certain success in quantitative comparison of theoretical
results with observations.

However, the recent observations~\cite{Ton} have shown that $w \leq -1$ (see Ref.~\cite{Star1} where this phenomenon
is described by a model-independent way).  Such a
type of DE, also known as phantom energy~\cite{Cald}, can be introduced
within the framework of both directions mentioned above. In
particular, in case of scalar fields phantom energy is realized by
change of a sign before a kinetic term in a field Lagrangian.
Choosing, further, a certain form of potential energy of this
field, it is possible to get a satisfactory agreement with
observations by selecting parameters of a model (see, e.g., \cite{Capoz1} and
references therein). On the other hand, one can try to model an inflation
in the early Universe using phantom fields~\cite{Piao} and consider quantum evolution of
the early Universe within the framework of phantom cosmology~\cite{Kiefer}.

Phantom DE for two scalars (where only one of scalars is phantom)
was considered in \cite{Elizalde:2004mq}. The idea of unification of
inflation with acceleration for phantoms was proposed in
\cite{Nojiri:2003vn}. The paper \cite{Szydlowski:2004jv} is devoted
to the analysis of conformally coupled phantom scalar fields. The
authors apply the Ziglin theory of integrability and find that the
flat model is nonintegrable. It means that we cannot expect to
determine simple analytical solutions of the Einstein equations. The
metric presented here shows the same: a solution of the Einstein
equations coupled with two interacting scalar fields can be found
only in a numerical form. General review on modified gravity as DE
one can find in \cite{Nojiri:2006ri}.

Another possible application of phantom scalar fields consists in modelling
of the so-called spacelike branes (S-branes) with their help. The S-branes are topological defects localized on a spacelike
hypersurface. An example of such a solution is a domain wall in a 4D scalar field
theory~\cite{Gut}. In this model the theory of one scalar field with effective potential energy with spontaneous
symmetry breaking is under consideration. This effective potential has two minima. For a static case, this problem
has the well-known solutions in the form of kinks~\cite{Raj}. These solutions start from one minimum (at $x\rightarrow -\infty$) and
tend asymptotically to another minimum of the effective potential (at $x\rightarrow +\infty$). However, if one
consider phantom fields, the similar solutions in a form of kink will be obtained not for the space variable
$x$ but for the time $t$. This so-called S2-brane solution is a timelike kink~\cite{Sen}.

In this paper we consider
a model of S-brane formed by two phantom fields having local and global minima.
Below we show that solutions similar to the soliton-like solutions mentioned above
could exist and for a case when an effective potential is choosing
in the form of two interacting scalar fields \eqref{pot}. This potential is an analog of the well-known model with
one scalar field with a Mexican hat potential. The fundamental difference of these two potentials consists in presence
of two scalar fields in \eqref{pot} that provide existence not only kink-like solutions but also another types
of soliton-like solutions.

\section{Equations}

Let us consider the flat cos\-molo\-gical Friedmann model with the
metric:
\begin{equation}
\label{metric}
ds^2=dt^2-a(t)^2 (dx^2+dy^2+dz^2).
\end{equation}
The Lagrangian for phantom scalar fields $\phi, \chi$ plus gravitation is:
\begin{equation}
L=-\frac{R}{16 \pi G}-\frac{1}{2}\partial_\mu \phi \partial^\mu
\phi-\frac{1}{2}\partial_\mu \chi \partial^\mu
\chi-V(\phi,\chi),
\end{equation}
where
\begin{equation}
\label{pot}
V(\phi,\chi)=\frac{\lambda_1}{4}(\phi^2-m_1^2)^2+\frac{\lambda_2}{4}(\chi^2-m_2^2)^2+\phi^2 \chi^2-V_0.
\end{equation}
Here $V_0$ is defining by initial conditions and can be considered as a cosmological constant $\Lambda$.
Introducing the dimensionless curvature tensor $\rho_i^k=l_{pl}^2 R_i^k$, the Einstein and scalar field equations
can be written as
\begin{equation}
\label{einst}
\rho_i^k-\frac{1}{2}\delta_i^k \rho=8\pi G l_{pl}^2 T_i^k,
\end{equation}
\begin{equation}
\label{ph}
\frac{1}{\sqrt{-g}}\partial_\mu(\sqrt{-g}g^{\mu\nu}\partial_\nu \phi)=\frac{\partial V(\phi,\chi)}{\partial \phi},
\end{equation}
\begin{equation}
\label{ch}
\frac{1}{\sqrt{-g}}\partial_\mu(\sqrt{-g}g^{\mu\nu}\partial_\nu \chi)=\frac{\partial V(\phi,\chi)}{\partial \chi},
\end{equation}
where $G$ is the gravitational constant, $l_{pl}$ - the Planck length. Using the metric \eqref{metric} and introducing dimensionless
variables and constants $\phi\rightarrow \phi/\sqrt{8\pi}l_{pl}$, $\chi\rightarrow \chi/\sqrt{8\pi}l_{pl}$,
$t\rightarrow \sqrt{4\pi}l_{pl} t$, $m_1 \rightarrow m_1/\sqrt{8\pi}l_{pl}$, $m_2 \rightarrow m_2/\sqrt{8\pi}l_{pl}$,
$\lambda_{1,2}\rightarrow 2\lambda_{1,2}$ one can rewrite the equations \eqref{einst}-\eqref{ch} as (in the Planck units, i.e.
at $c=\hbar=G=1$):
\begin{equation}
\label{einst0}
\left( \frac{\dot a}{a}\right)^2=\frac{1}{6}\left[-\dot \phi^2-\dot \chi^2+V(\phi,\chi)\right]
\end{equation}
\begin{equation}
\label{einst1}
\frac{\ddot a}{a}-\left (\frac{\dot a}{a}\right)^2=\frac{1}{4}\left ( \dot \phi^2+\dot \chi^2 \right),
\end{equation}
\begin{equation}
\label{ph1}
\ddot \phi +3\frac{\dot a}{a}\dot \phi=\phi \left[\chi^2+\lambda_1\left(\phi^2-m_1^2\right)\right],
\end{equation}
\begin{equation}
\label{ch1}
\ddot \chi +3\frac{\dot a}{a}\dot \chi=\chi \left[\phi^2+\lambda_2\left(\chi^2-m_2^2\right)\right],
\end{equation}
where
\begin{equation}
\label{V}
V(\phi,\chi)=\frac{\lambda_1}{2}(\phi^2-m_1^2)^2+\frac{\lambda_2}{2}(\chi^2-m_2^2)^2+\phi^2 \chi^2-V_0,
\end{equation}
and
\begin{equation}
\label{V0}
V_0=\frac{\lambda_1}{2}(\phi_0^2-m_1^2)^2+\frac{\lambda_2}{2}(\chi_0^2-m_2^2)^2+\phi_0^2 \chi_0^2.
\end{equation}
The last value follows from Eqs. \eqref{einst0}, \eqref{V} and the
initial conditions \eqref{in1}-\eqref{in3}. (The equation
\eqref{einst1} was obtained by subtraction of the $\left(^t_t\right)$ component
of \eqref{einst} from the $\left(^x_x\right)$ component.) For these equations, there
are following initial conditions:
\begin{eqnarray}
a(0)=a_0, \quad \dot a(0)=0,
\label{in1} \\
\phi(0)=\phi_0, \quad \dot \phi(0)=0,
\label{in2}\\
\chi(0)=\chi_0, \quad \dot \chi(0)=0.
\label{in3}
\end{eqnarray}

In Ref. \cite{Dzhun1} the research of mathematically similar equations have been carried out.
Those equations describe the thick brane in the Randall-Sundrum scenario.
 It was shown that at choice of the potential in the form \eqref{V},
there are the regular solutions describing the thick brane. Such
solutions can be obtained only if the potential has local and global
minima. In the context of Ref. \cite{Dzhun1}, the specified scalar fields could be interpreted as a condensate of a
gauge field.

Having a purpose to obtain regular cosmological solutions,
we choose phantom fields with the given potential as a model of scalar field.
In this case Eqs.  \eqref{einst0}-\eqref{ch1} are mathematically similar to the equations describing the thick brane
in  Ref. \cite{Dzhun1}.

\section{Numerical analysis}

For the numerical calculations we choose the following values of the parameters and initial conditions:
\begin{equation}
a_0=1, \quad \phi_0=\sqrt{3}, \quad \chi_0=\sqrt{1.8}, \quad \lambda_1=0.1, \quad \lambda_2=1.0.
\end{equation}
We apply the methods of step by step approximation for finding of numerical solutions (the details of similar
calculations can be found in Ref. \cite{Dzhun1}).

{\bf Step~1.} On the first step we solve Eq. \eqref{ph1} (having zero approximations $a_0(t)=a_{0} e^{t}$,
$\chi_0(t)=\sqrt{0.6}/\cosh^2(t/4)$).
The regular solution exists for a special value $m^*_{1,i}$ only. For $m_1<m^*_{1,i}$ the function $\phi_i(t)\rightarrow +\infty$
and for $m_1>m^*_{1,i}$ the function $\phi_i(t)\rightarrow -\infty$ (here the index $i$ is the approximation number).
One can say that in this case we solve {\it a non-linear eigenvalue problem}: $\phi_i^*(t)$ is the eigenstate and $m^*_{1,i}$
is the eigenvalue on this Step.

{\bf Step~2.} On the second step we solve Eq. \eqref{ch1} using zero approximation $a_0(t)$ for the function $a(t)$ and the
first approximation $\phi_1^*(t)$ for the function $\phi(t)$ from the Step 1. For $m_2<m^*_{2,1}$ the function
$\chi_1(t)\rightarrow +\infty$ and for $m_2>m^*_{2,1}$ the function $\chi_1(t)\rightarrow -\infty$.
 Again we have {\it a non-linear eigenvalue problem} for the function $\chi_1^*(t)$ and $m^*_{2,1}$.

{\bf Step~3.} On the third step we repeat the first two steps that to have the good convergent sequence
$\phi_i^*(t), \chi_i^*(t)$. Practically we have made three approximations.

{\bf Step~4.} On the next step we solve Eq. \eqref{einst1} which gives us the function $a_1(t)$.

{\bf Step~5.} On this step we repeat Steps 1-4 necessary number of times that to have the necessary accuracy of definition
of the functions $a^*(t), \phi^*(t), \chi^*(t)$ and the eigenvalues $m_{1,2}^*$.

\begin{figure}[h]
\begin{minipage}[t]{.49\linewidth}
  \begin{center}
  \includegraphics[width=8cm]{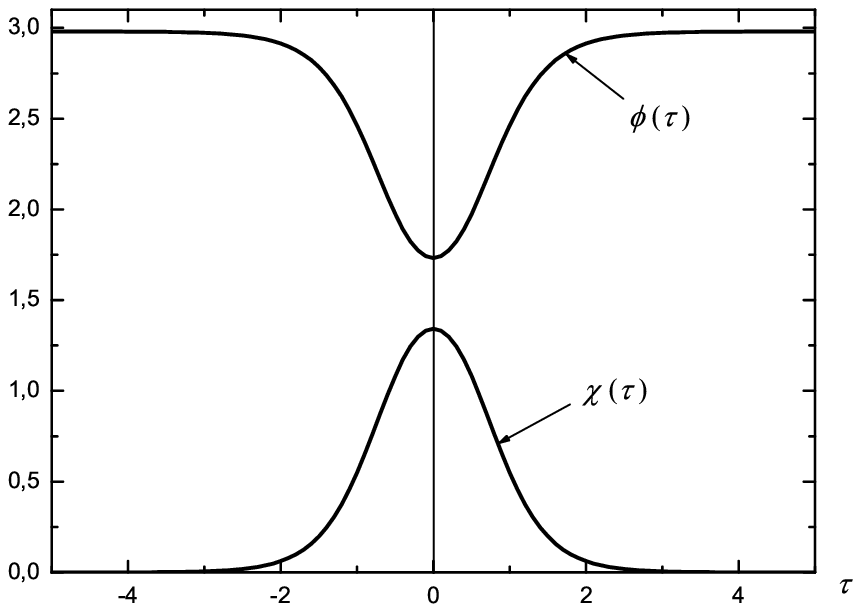}
  \caption{The graphs for the scalar fields $\phi^*(t), \chi^*(t)$. Both functions tend
  asymptotically to the local minimum at $\phi^*(t)=m_1^*, \chi^*(t)=0$.  }
    \label{phch}
  \end{center}
\end{minipage}\hfill
\begin{minipage}[t]{.49\linewidth}
  \begin{center}
  \includegraphics[width=8cm]{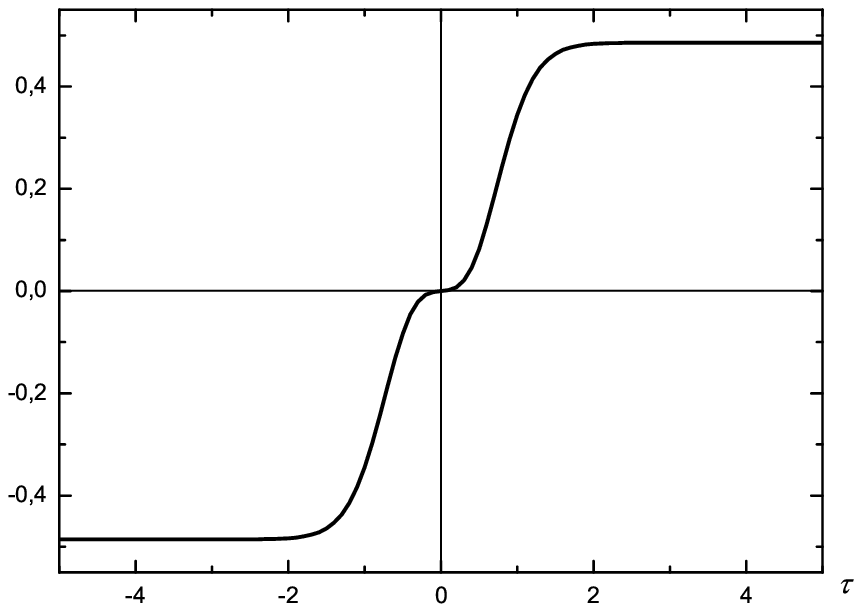}
  \caption{The Hubble parameter $H=\dot a/a$ tends asymptotically to
  the de Sitter stages: if $H<0$ - the stage of contraction, if $H>0$ - the
  stage of expansion.}
    \label{fig_h}
  \end{center}
\end{minipage}\hfill
\end{figure}

After Step~5 we have the solution presented in Figs. \eqref{phch}, \eqref{fig_h}. These numerical calculations give us the eigenvalues
$m_1^*=2.98530175, m_2^*=2.538974725$ and eigenstates $a^*(t), \phi^*(t), \chi^*(t)$.

\section{Conclusions}

The toy model of inflation in the early Universe with account of the
phantom fields is considered. The possibility of obtaining of bounce
solutions is shown in principle. In this case the model evolves from
a stage of contraction ($\dot a<0$), passes through a minimum of the
scale factor ($a_0\sim 1$ in the Planck units) and appears on the de
Sitter stage of expansion (Fig. \eqref{fig_h}). Such evolution is
provided in the model by a special form of the potential for the phantom fields
\eqref{V}.

From the mathematical point of view, the problem was reduced to
finding of eigenfunctions of the nonlinear differential equations
\eqref{einst1}-\eqref{ch1}. Only at the obtained values of $m_1^*,
m_2^*$, the solutions are regular and a bounce and an inflation take
place. At other values of these parameters, the solutions are
singular.

Let us note that at the obtained values of the parameters $m_{1,2}=m_{1,2}^*$, the de Sitter stage
occurs over an infinite period of time. But if $|m_{1,2}-m_{1,2}^*|\ll 1$ than there is a finite time of
inflation of the Universe.

The obtained solutions could be interpreted as S2-brane solutions for a case of two interacting  phantom
scalar fields. But in contrast to the usual kink-like S2-brane solutions~\cite{Gut, Sen},
our solutions are another type of soliton-like solutions. If the kink solution starts from one minimum
of a potential and tends asymptotically to another one, our solutions start and finish in the same minimum
(in the case under consideration, in the local minimum $\phi=m_1, \chi=0$, see Fig.~\eqref{phch}). By the
terminology of~\cite{Raj}, the kink-like solutions refer to topological solutions and our solutions - to
non-topological ones. In contrast to the kink-like solutions which exist only in a case of presence of two
minima, the existence of non-topological solutions is also possible at one minimum only.

\end{document}